\documentclass[12pt]{article}

\def\rdots{\mathinner{\mkern1mu\raise1pt\vbox{\kern1pt\hbox{.}}\mkern2mu
   \raise4pt\hbox{.}\mkern2mu\raise7pt\hbox{.}\mkern1mu}}
\newcommand{\Z}{{\rm Z\kern-.35em Z}}
\newcommand{\bP}{{\rm I\kern-.15em P}}
\newcommand{\Q}{\kern.3em\rule{.07em}{.65em}\kern-.3em{\rm Q}}
\newcommand{\R}{{\rm I\kern-.15em R}}
\newcommand{\h}{{\rm I\kern-.15em H}}
\newcommand{\C}{\kern.3em\rule{.07em}{.65em}\kern-.3em{\rm C}}
\newcommand{\T}{{\rm T\kern-.35em T}}

\newcommand{\be}{\begin{equation}}
\newcommand{\ee}{\end{equation}}

\newcommand{\pa}{\partial}

\newcommand{\ra}{\rightarrow}

\newcommand{\al}{\alpha}
\newcommand{\nn}{\nonumber}

\begin{document}
\font\cs=cmssi12

\openup 1.5\jot
\centerline{A Polymer Expansion for the Random Walk on the Permutation Group}
\centerline{Associated to the Quantum Heisenberg Ferromagnet}

\vspace{1in}
\centerline{Paul Federbush}
\centerline{Department of Mathematics}
\centerline{University of Michigan}
\centerline{Ann Arbor, MI 48109-1109}
\centerline{(pfed@umich.edu)}

\vspace{1in}

\centerline{\underline{Abstract}}

For a long time one has associated to the Quantum Heisenberg Ferromagnet on a lattice, a random walk on the permutation group of the lattice vertices.  We here present a polymer expansion for the solution of the heat equation coupled to the random walk.  We work on a finite lattice, there is no question of convergence.  We leave to future work bounding terms in the expansion necessary to extend the result to an infinite lattice.
\vfill\eject

The random walk on the permutation group is discussed in [3] and [4].  The polymer expansion which we will present is parallel to the development in [5] of a polymer expansion for the wave function of the Heisenberg ferromagnet.  One should read that paper, [5], as though it were a warm up example.  We find the current expansion as interesting in its own right, whether or not it ever is used in say a proof of the phase transition for the magnet.  The bounding of polymer contributions, and the extension of the results to an infinite lattice, both left to future work, seem to present real mathematical challenges in novel directions.

We work with a finite rectangular lattice, $V$, in $d$-dimensions, $\cal V$ the set of its vertices.  $G$ is the permutation group on the elements of $\cal V$.  We use the notation
\be
\vec{\cal S} = \left({\cal S}^{(1)}, \  {\cal S}^{(2)} \right)
\ee
where ${\cal S}^{(1)}$ and $ {\cal S}^{(2)}$ are both subsets of $\cal V$, of the same cardinality.  We let
\be
\vec{\cal V} = ({\cal V}, {\cal V})
\ee
By a partition of $\vec{\cal S} $, with elements $\vec{\cal S}_\al < \vec{\cal P}$ we require
\begin{eqnarray}
\cup \ {\cal S}_\alpha^{(1)} &=& {\cal S}^{(1)} \\
\cup \ {\cal S}_\alpha^{(2)} &=& {\cal S}^{(2)} \nonumber \\
\# \big( {\cal S}_\alpha^{(1)} \big) &=& \# \big({\cal S}^{(2)}_\al \big) \nn \\
{\cal S}_\alpha^{(1)} \cap  {\cal S}^{(1)}_\beta &=& {\cal S}_\alpha^{(2)} \cap  {\cal S}^{(2)}_\beta = \emptyset \ {\rm if} \ \al \not= \beta. \nn
\end{eqnarray}

We also work with ordered sets and subsets.  We write
\be  {\vec{\bf o}} = \left( {\bf o}^{(1)}, \  {\bf o}^{(2)} \right)   \ee
where $ {\bf o}^{(1)}$ and  ${\bf o}^{(2)}$ are each ordered subsets of $\cal V$, of the same cardinality.  We put an arbitrary but fixed "standard" ordering on $\cal V$, and assume the ordering of ${\bf o}^{(1)}$ is always that inherited from the standard ordering of $\cal V$.  We write
\begin{eqnarray}
\vec{\bf o} \rightarrow \vec{\cal S} & & \ \ {\rm if} \nn \\
{\bf o}^{(1)}& & \ \ {\rm is\;an\;ordering\;of} \ \  {\cal S}^{(1)} \ \ {\rm and} \\
 {\bf o}^{(2)}& & \ \ {\rm is\;an\;ordering\;of} \ \  {\cal S}^{(2)}. \nn
\end{eqnarray}
We also write
\begin{eqnarray}
\vec{\bf o} \rightarrow {\cal S} & & \ \ {\rm if} \\
 {\bf o}^{(1)}& & \ \ {\rm is\;an\;ordering\;of} \ \  {\cal S}. \nn
\end{eqnarray}
We notice that elements, $g$, of $G$ are in 1 - 1 correspondence with $\vec{\bf o}$ such that $\vec{\bf o} \ra \vec{\cal V}$ (or equivalently $\vec{\bf o} \ra {\cal V}$) , by
\be       g : {\bf o}^{(1)}(i) \ra {\bf o}^{(2)}(i), \ \ \ \ \  i=1,..., \# ({\cal V}).		\ee
where ${\bf o}(i)$  is ith element of ordered set ${\bf o}$.

We have a function of $t$ on the permutation group, $C(\vec {\bf o}, t)$, where $\vec {\bf o} \ra {\cal V}$, and we are identifying $\vec {\bf o}$ with a group element by (7).  We will also write $C(\vec {\bf o}, t) = c^{\cal V}(\vec {\bf o},t)$.  $C(\vec {\bf o}, t)$ satisfies the heat equation that also defines the random walk on the permutation group, [3], [4].
\be
\frac{\pa C(\vec {\bf o},t)}{\pa t} = \sum_{\vec {\bf o}' \sim \vec {\bf o}} \left( C(\vec {\bf o}', t) - C(\vec {\bf o},t)\right).\ee
Here $\vec {\bf o} \ra \cal V$ and $\vec {\bf o}' \ra \cal V$, and $\vec {\bf o}' \sim \vec {\bf o}$,  if $ {\bf o}^{(2)}$ and ${\bf o}'^{(2)}$ differ exactly by the interchange of two vertices, vertices that are nearest neighbors on the  lattice, $ V$.

If $\cal S$ is a subset of $\cal V$, we define $c^{\cal S}(\vec {\bf o},t)$, for $\vec {\bf o} \ra \cal S$.  For two ordered sets $\vec{\bf o}$ and $\vec{\bf s}$ we say
\be	\vec{\bf o} \subset \vec{\bf s}   \ee
if there are positive integers $i_1 < i_2 < \cdots < i_s$ such that
\[	{\bf o}^{(1)} = \left\{ {\bf s}^{(1)}(i_1), \cdots, {\bf s}^{(1)}(i_s) \right\}   \]
\be \ee
\[	{\bf o}^{(2)} = \left\{ {\bf s}^{(2)}(i_1), \cdots, {\bf s}^{(2)}(i_s) \right\}   \]
each here ordered sets.  Then we set
\be	c^{\cal S}(\vec {\bf o},t) = 
\begin{array}[t]{c}
{\displaystyle\sum_{\vec{\bf s}}} \\
{\scriptstyle \vec{\bf o} \subset \vec{\bf s}}
\end{array}
C(\vec{\bf s}, t)
\ee
the sum in (11) of ordered sets $\vec{\bf s}$, with $\vec{\bf s} \ra \cal V$.  These also satisfy heat equations
\be \frac \pa{\pa t} c^{\cal S}(\vec {\bf o},t) = \sum_{\vec {\bf o}' \sim \vec {\bf o}} \left( c^{{\cal S}}(\vec {\bf o}', t) - c^{{\cal S}}(\vec {\bf o},t)\right).
\ee
These equations are as in [4].  In (12) we have set, for $\vec {\bf o}$ and $\vec {\bf o}' $ satisfying $\vec {\bf o} \ra {\cal S}, \ \vec {\bf o}' \ra \cal S$
\be	\vec {\bf o}'  \sim \vec {\bf o}	\ee
if either
\begin{itemize}
\item[(1)] ${\bf o}^{(2)}$ and  ${\bf o}'^{(2)}$ differ exactly by the interchange of two vertices, vertices that are nearest neighbors on the lattice, $V$ 

or

\item [(2)]  ${\bf o}'^{(2)}$ is obtained from ${\bf o}^{(2)}$ by replacing one of the vertices in ${\bf o}^{(2)}$ by one of its nearest neighbors on lattice , $V$ (that must of course not be one of the other vertices in ${\bf o}^{(2)}$.)
\end{itemize}

The polymer expansion for $C(\vec {\bf o},t)$ is now presented, in terms of quantities to be further explicated.
\be
C(\vec{\bf o}, t) = \sum_{\vec{\cal P}} \left( \prod_{ \vec{\cal S}_\al < \; \vec{\cal P} } \left(
\begin{array}[t]{c}
{\displaystyle\sum_{\vec{\bf o}_k \ra \vec{\cal S}_\al}} \\
{\scriptstyle \vec{\bf o}_k \subset \vec{\bf o}}
\end{array}
u(\vec{\bf o}_k, t) \right) \right)
\ee
where here $\vec{\cal P}$ is a partition of $\vec{\cal V}$.  The restriction in the last sum, $\vec{\bf o}_k \subset \vec{\bf o}$, is of course necessary; but perhaps one is disappointed that this equation (14) does not look more like equation (13) of [5].  We must recognize though that the system treated here is rather different than the Schroedinger equation of [4] for the magnet wave function, and one should be pleased there is as much similarity as there is.  We proceed to study the $u$'s.

We wish to find an expression for $c^{{\cal S}}(\vec {\bf o},t)$ defined in (11) from the expression for $C(\vec {\bf o},t)$ in (14).  We first find it convenient  to define a specialized type of partition.  $\vec {\cal P}$.  A partition of $\vec {\cal V}$, is ``${\cal S}$-covering" if it consists exactly of subsets $\{ \vec{\cal S}_\al \}$ and $\vec{\cal S}^c$ that satisfy 
$\{ {\cal S}^{(1)}_\al \cap {\cal S} \}$ is a partition of $\cal S$.  (We never accept the empty set as an element of a partition.)

We also must define the restriction of $\vec{\bf o}$ to a set $\cal S$, $\vec{\bf o}|_{\cal S}$, by in viewing $\vec{\bf o}$ as the ordered set of pairs $\Big({\bf o}^{(1)}(i), \; {\bf o}^{(2)}(i)\Big),  \ \  i=1,..., \# ( {\bf o}^{(1)})$, (see remark after equation (7)), letting $\vec{\bf o}|_{\cal S}$ be constructed by throwing away pairs where ${\bf o}^{(1)}(i) \not\in \cal S$, and keeping same ordering on remaining pairs.

We find then:
\be
c^{{\cal S}}(\vec{\bf o}, t) = \sum_{\vec{\cal P}} \left( \prod_{\vec{\cal S}_\al < \vec{\cal P}} \left(
\begin{array}[t]{c}
{\displaystyle\sum_{\vec{\bf o}_k \ra \vec{\cal S}_\al}} \\
{\scriptstyle \vec{\bf o}_k|_{{\cal S}} \subset \vec{\bf o}}
\end{array}
u(\vec{\bf o}_k, t) \right) \right) w(\vec{\cal S}^c, t)
\ee
where $\vec{\bf o} \ra \cal S$, the $\vec{\cal P}$ are $\cal S$-covering partitions of $\vec{\cal V}$, and $w(\vec{\cal S}^c, t)$ and $u(\vec{\bf o}_k, t)$ remain to be specified.

The expression for $w$ is given in terms of the $u$'s.
\be
w(\vec{\cal S}, t) = \sum_{\vec{\cal P}} \left( \prod_{\vec{\cal S}_\al < \vec{\cal P}} \left(
\sum_{\vec{\bf o}_k \ra \vec{\cal S}_\al} u(\vec{\bf o}_k, t) \right) \right)
\ee
Here $\vec{\cal P}$ is a partition of $\vec{\cal S}$.  For the polymer expansion of [5] one obtained a unique expression for the corresponding $u$'s.  Here the $u$'s and $w$'s are any quantities satisfying the sets of equations (14), (15), and (16).  We have considered several conditions that might be imposed to limit solutions.  An attractive condition might be the imposition of 
\[	\sum_{\vec{\bf o}_k \ra \vec{\cal S}} u(\vec{\bf o}_k, t) = 0 \ \ \ {\rm if} \ \ \  \#({\cal S}^{(1)}) > 1.    \]
There are certainly solutions, since there is always the ``trivial" solution where all $u$'s are zero except the $\vec u(\vec{\bf o}_k, t)$ with $\vec{\bf o}_k \ra \cal V$.  We would of course be interested in solutions where the $u$'s are constructed in some manner inductively over the ascending $\#({\bf o}^{(1)}_k)$.  At the initial inductive step we would have, with $\#({\bf o}^{(1)}) = 1$
\begin{eqnarray}
u(\vec{\bf o}, t) &=& u\Big( ({\bf o}^{(1)}, {\bf o}^{(2)}), t\Big) = u \Big((\{i\}, \{j\}), t\Big) = c g(t)_{ij} \\
w(\{i\}^c, t) &=& 1
\end{eqnarray}
and other $u$'s set zero.  $g(t)_{ij}$ is the lattice heat equation Green's function.  In $\{ i\}^c$ $c$ indicates complement.   We find determining the constant $c$ in (17) a challenging problem.  We leave study of the rest of an inductive construction of the $u$'s to further work.

We cannot resist presenting a hand waving computation of the $c$ of (17).  We want $C(\vec{\bf o}, 0)$ to be 1 if $\vec{\bf o} = (\hat{\cal V}, \hat{\cal V})$, where $\hat{\cal V}$ is $\cal V$ in the standard ordering, and zero for other $\vec{\bf o}$.  If $C(\vec{\bf o},t)$ is constructed from exactly the $u$'s of equation (17), then one must have
\be 1 = \sum_{\it p} \prod_i \left(cg(t)_{ip(i)} \right) \ee
where {\it p} is a permutation of $1, 2, \cdots, \#(\cal V)$.  We assume
\be	\sum_i g_{ij}(t) = 1.	\ee
We now assume the points in ${\cal V}^{\#({\cal V})}$ given by $(p(i))_i$ as $p$ varies are ``random".  One would have by a monte-carlo argument
\be	1 = \frac {N!}{N^N} \ c^N	\ee
where $N = \#(\cal V)$.  So we get
\be	c \cong e \ .	\ee
We find this argument amusing, and believe in some form (21) becomes true as $t$ gets large.

We conclude with two interesting theorems related to the current study.  They are striking enough to include just for amusement.  But also they can possibly be used to coordinate computation of the $u$'s and find further relations for these quantities.

We go from the $c^{{\cal S}}(\vec{\bf o}, t)$ that are solutions to a heat equation on the permutation group, to solutions of the heat equation in fixed number spin wave states of the Heisenberg magnet.  (One could probably be very abstract, working with two ``categories" and a ``forgetful functor" from one to the other....we will be very concrete instead.)  We define
\be	\tilde c^{{\it s}}(s',t) = \sum_{\vec{\bf o} \ra ({\it s},{\it s}')} c^{{\it s}}(\vec{\bf o}, t)	\ee
where ${\it s}'$ and ${\it s}$ must have same cardinality.

\bigskip

\noindent
\underline{The Forgetful Theorem} $\tilde c^{{\it s}}({\it s}',t)$ satisfies the heat equation of $\#({\it s})$ spin wave state of the magnet.  That is, equation (10) of [3] (replace ${\it s}$ and ${\it s}'$ of equation (10) of [3] by ${\it s}'$ and ${\it s}''$, and $r$ by $\#( {\it s}$)).

\bigskip

Using ${\cal S}^*$ for complement of ${\cal S}$ in ${\cal V}$ one also has 

\noindent
\underline{The Duality Theorem}.
\be   \tilde c^{{\it s}}({\it s}',t) = \tilde c^{{\it s}^*}({\it s}'^*, t)    .    \ee

\bigskip

The duality is that between spin wave states of complementary number.

\vfill\eject 

\centerline{\underline{References}}

\begin{itemize}
\item[[1]] P. Federbush, ``For the Quantum Heisenberg Ferromagnet, Some Conjectured Approximations", math-ph/0101017.
\item[[2]]  P. Federbush, ``For the Quantum Heisenberg Ferromagnet, A Polymer Expansion and its High T Convergence", math-ph/0108002.
\item[[3]] P. Federbush, ``For the Quantum Heisenberg Ferromagnet, Tao to the Proof of a Phase Transition", math-ph/202044.
\item[[4]] Robert T. Powers, ``Heisenberg Model and a Random Walk on the Permutation Group", {\it Lett. in Math. Phys.} {\bf 1}, 125-130 (1976).
\item[[5]] P. Federbush, ``A Polymer Expansion for the Quantum Heisenberg Ferromagnet Wave Function", math-ph/0302067.

\end{itemize}

\end{document}